\documentclass[twocolumn,preprintnumbers,amsmath,amssymb]{revtex4}

\usepackage{graphicx}
\usepackage{dcolumn}
\usepackage{bm}

\begin{document}


\title{
ON SATURATION OF CHARHED HADRON PRODUCTION \\[1mm]
IN PP COLLISIONS AT LHC\\[3mm]
}

\author{M.V.~Tokarev$^{\star}$}
\author{I.~Zborovsk\'{y}$^{\ddagger}$}
\affiliation{$^{\star}$Joint Institute for Nuclear Research, Dubna, Russia
\\
 $^{\ddagger}$Nuclear Physics Institute,
Academy of Sciences of the Czech Republic, \\
\v {R}e\v {z}, Czech Republic}
\begin{abstract}
First results on charged hadron transverse momentum spectra in
$pp$ collisions obtained  by the CMS Collaboration at LHC were
analyzed in $z$-scaling approach.
The first LHC data confirm $z$-scaling.
The saturation regime of the
scaling function $\psi(z)$
observed in $pp$ and $\bar pp$ interactions at lower energy
$\sqrt s = 19-1960$~GeV  is verified.
The saturation of $\psi(z)$  for charged hadrons is found down to $z\simeq 0.05$
at the highest energy $\sqrt s=2360$~GeV reached till now at colliders.
A microscopic scenario of hadron production is discussed
in connection with search for new signatures of phase transitions
in hadron matter.
Constituent energy loss
and its dependencies on the transverse momentum
and collision energy are estimated.
The beam energy scan at LHC in the saturation region
is suggested.
\\[5mm]

\end{abstract}

\maketitle

\section{Introduction}

Commissioning of the Large Hadron Collider (LHC) at CERN
aimed to discover new physics beyond Standard Model (SM)
gives a unique possibility \cite{CMS1,CMS2,CMS3,ATLAS1,ATLAS2, ALICE1, ALICE2}
to verify various theoretical models
and experimental properties of hadron interations
established at ISR, S$\bar{\rm p}$pS, SPS, RHIC, and Tevatron
over the range $\sqrt s =19-1960$~GeV
\cite{ISR_pp1,ISR_pp2,Cronin_pp,Jaffe_pp,STAR_pp}.
Phenomenological features of particle production experimentally
found and theoretically predicted
are extremely important for scientific search
on high energy physics frontier
\cite{Bjorken,Feynman,Pol,KNO,Matveev,Brodsky,Bialas,Dremin,
DeWolf}.

One of such features is a new scaling  ($z$-scaling) of hadron production
in $pp$ and $\bar pp$ collisions suggested
in \cite{Z1} (see also {\cite{Z2,PRD75,IJMP24} and references therein).
It was used as a method allowing systematic analysis of data on
inclusive cross sections and search for new physics.
The scaling function $\psi(z)$
and scaling variable $z$
are expressed via experimentally measurable quantities.
The shape of the scaling function was shown to be independent
of the energy, multiplicity density of collisions,
detection angle and hadron type including
production of particles with heavy flavor content.
The power behavior $\psi(z)\sim z^{-\beta}$
was established in the high-$z$ range.
At low $z$, a saturation \cite{PRD75,IJMP24}
of the scaling function
was found down to $z\simeq 10^{-3}$.
The single parameter
$c$ which controls the behavior of $\psi(z)$ at low $z$ is
interpreted as a "specific heat" of the produced
medium associated with inclusive particle.
The scaling
in $pp$ and $p\bar p$ collisions is consistent with a constant
value of $c$. Possible change in this parameter
is assumed to be an indication of a phase transition of the
matter produced in high energy collisions.
Investigation of the non-perturbative regime and phase transitions
in non-Abelian theories is considered to be preferred in
the soft region (low-$p_T$ and high multiplicity) of hadron production.
It was concluded that $z$-scaling reflects self-similarity
of particle structure, interaction of constituents, and hadronization process.

In the present paper we analyze the first data \cite{CMS_pp}
on transverse momentum spectra of charged hadrons produced in $pp$ collisions
at the energy $\sqrt s = 900$ and 2360~GeV in the middle rapidity range
obtained by the CMS Collaboration at LHC.
The saturation of $\psi(z)$ at the LHC energies is verified.
The results are compared with ISR, S$\bar{\rm p}$pS,
SPS,
RHIC, and Tevatron data.
The microscopic scenario of hadron production in the $z$-scaling approach
is used to estimate the energy loss and recoil mass at the constituent level
in dependence of the collision energy and transverse momentum of the inclusive hadron.
The change of collision and particle reconstruction conditions
(energy, multiplicity, type of particle, transverse momentum)
is discussed in connection with possibility for a Beam Energy Scan (BES)
program at the LHC to expand research area in search of strongly pronounced
signatures of phase transitions.
We consider that systematic analysis of particle
production as a function of the collision energy and multiplicity
towards searching for clear signatures of phase transition
from quark and gluon to hadron degrees of freedom
as well as location of the Critical Point (CP) on the QCD phase diagram is possible
at energies achieved at the LHC now.

\section{$z$-Scaling}

Here we briefly remind the basic ideas
of the $z$-scaling concept \cite{PRD75,IJMP24}.
The collision of extended objects (hadrons, nuclei)
at sufficiently high energies
is considered as an ensemble of individual interactions
of their constituents (partons, quarks, gluons).
A single interaction of the constituents is illustrated in Fig.1.
\begin{figure}
\begin{center}
\vskip 5mm
\hspace*{0mm}
\includegraphics[width=55mm,height=55mm]{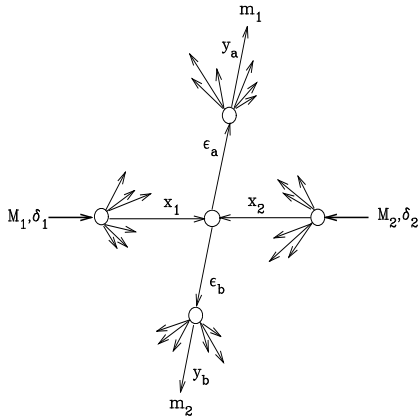}
\caption{Diagram of the constituent subprocess
of the reaction $p+p\rightarrow h+X$.}
\end{center}
\end{figure}
Structures of the colliding objects are characterized by the
parameters $\delta_1$ and $\delta_2$. The constituents of the
incoming objects (hadrons or nuclei) with the masses $M_1, M_2$ and momenta
$P_1, P_2$ carry their fractions $x_1, x_2$. The inclusive
particle carries the momentum fraction $y_a$ of the
object produced in the constituent collision into the observed direction.
Its fragmentation is characterized by a parameter
$\epsilon_a$. The fragmentation in the recoil direction is
described by a parameter $\epsilon_b$ and the momentum fraction $y_b$.
Multiple interactions of the constituents are considered to be similar.
This property reflects a self-similarity of the hadronic interactions
at the constituent level.

\subsection{Momentum fractions $x_1, x_2, y_a$, and $y_b$}

The elementary subprocess is considered to be a binary collision
of the constituents $(x_{1}M_{1})$ and $(x_{2}M_{2})$ resulting in
the scattered $(m_1/y_a)$ and recoil $(x_{1}M_{1}+x_{2}M_{2} +
m_2/y_b)$ objects in the final state.
The produced secondary objects transform into real particles
after the constituent collisions.
The registered particle with the mass $m_1$ and the 4-momentum $p$
is produced with
its hadron counterpart $(m_2)$ carrying
the momentum fractions $y_b$ of the produced recoil.
The momentum conservation law of the constituent subprocess
is written in the form
\begin{equation}
(x_1P_1+x_2P_2-p/y_a)^2 = M_X^2,
\label{eq:r3}
\end{equation}
with the recoil mass $M_X=x_1M_1+x_2M_2+m_2/y_b$ .
The associate production of $(m_2)$ ensures conservation
of the additive quantum numbers.
Equation (\ref{eq:r3}) is an expression of the locality of the hadron interaction
at a constituent level. It represents a kinematical constraint on the momentum
fractions $x_1$, $x_2$, $y_a$, and $y_b$ which determine the underlying
elementary subprocess.

The structural parameters $\delta_1,\delta_2$ and $\epsilon_a,\epsilon_b$
are connected with the corresponding
momentum fractions by the function
\begin{equation}
\Omega =
(1-x_1)^{\delta_1}(1-x_2)^{\delta_2}(1-y_a)^{\epsilon_a}(1-y_b)^{\epsilon_b}.
\label{eq:r4}
\end{equation}
The quantity $\Omega$ is proportional to
relative number of all such constituent configurations
in the inclusive reaction
which contain the configuration defined by the fractions $x_1,x_2,y_a,$ and $y_b$.
The $\Omega$ plays the role of a relative volume
which occupy these configurations
in the space of the momentum fractions.
The parameters  $\delta_1,\delta_2$ and $\epsilon_a,\epsilon_b$
are interpreted as fractal dimensions in the
parts of the space
of the momentum fractions which correspond to the
colliding objects and fragmentation processes,
respectively.
For given values of $\delta_{1,2}$ and $\epsilon_{a,b}$,
the fractions $x_1$, $x_2$, $y_a$, and $y_b$
are determined in a way to maximize the function
$\Omega$, simultaneously fulfilling the condition (\ref{eq:r3}).

In the case of proton-proton interactions
we set $\delta_1=\delta_2\equiv\delta$.
We assume that the fragmentation
of the objects moving in the scattered and recoil
directions can be described by the same parameter
$\epsilon_a=\epsilon_b\equiv\epsilon$ which depends
on the type of the inclusive particle.
Values of the parameters $\delta$ and
$\epsilon$ are determined in accordance with the self-similarity
requirements and experiment.
They were found to have constant values
in $pp$  and $\bar pp$ collisions at high energies.

\vskip 5mm

\subsection{Scaling variable~$z$ and scaling function~$\psi(z)$}

The self-similarity of hadron interactions reflects a property
that hadron constituents and their interactions are similar.
The self-similarity variable $z$ is defined as follows
\begin{equation}
z =z_0 \Omega^{-1},
\label{eq:r8}
\end{equation}
where
\begin{equation}
z_0 = \frac{  \sqrt {s_{\bot}}    }{(dN_{ch}/d\eta|_0)^c  m}
\label{eq:r9}
\end{equation}
and $\Omega$ is maximal value of (\ref{eq:r4}) with the condition (\ref{eq:r3}).
For a given inclusive reaction the quantity $z$ is proportional to
the transverse kinetic energy $\sqrt {s_{\bot}}$ of the constituent subprocess
consumed on the production of the inclusive particle ($m_1$) and
its counterpart ($m_2$).
The quantity $dN_{ch}/d\eta|_0$ is
the corresponding multiplicity density
of charged particles in the central region of the inclusive reaction
at pseudorapidity $\eta=0$.
The parameter $c$ characterizes
properties of the produced medium and is interpreted as a "specific heat".
The mass constant $m$ is arbitrary and we fix it
at the value of nucleon mass.

The scaling function $\psi(z)$ is expressed in terms of the experimentally
measured inclusive cross section $Ed^3\sigma/dp^3$, the multiplicity
density $dN/d\eta$, and the total inelastic cross section $\sigma_{inel}$ as follows \cite{PRD75}
\begin{equation}
\psi(z) = -{ { \pi s} \over { (dN/d\eta) \sigma_{inel}} } J^{-1} E {
{d^3\sigma} \over {dp^3}},
\label{eq:r13}
\end{equation}
where $s$ is the square of
the center-of-mass energy and $J$ is the corresponding Jacobian.
The multiplicity density $dN/d\eta$ in (\ref{eq:r13})
depends on the center-of-mass energy, centrality, and
on the production angles at which the inclusive spectra were measured.
The function $\psi(z)$ is normalized to unity.
It is  interpreted  as a probability
density to produce an inclusive particle
with the corresponding value of the variable $z$.

\section{Self-similarity of hadron production}

Main features of the $z$-scaling in $pp$ interactions at FNAL, CERN,
and BNL energies were presented and discussed in \cite{PRD75,IJMP24}.
The experimental data cover
a wide range of the collision energy, transverse momenta, and angles
of the produced particles.
The energy, angular, and multiplicity independence
of the scaling function $\psi(z)$ gives strong constrains
on the values of the parameters $c$, $\delta$, and $\epsilon$.
The scaling is consistent with $c=0.25$ and $\delta=0.5$
for all types of the analyzed inclusive hadrons.
The value of $\epsilon$ increases with hadron mass.
The behavior of $\psi(z)$ is found to be described
by the power law,
$\psi(z)\sim z^{-\beta}$,
in the asymptotic high-$z$ (high-$p_T$) region \cite{PRD75}.
At low-$z$ (low-$p_T$), the scaling function flattens out and becomes a constant
for $z<0.1$ \cite{IJMP24}.

A typical example of $z$-presentation of the inclusive spectra
is shown in Fig.2.
\begin{figure}
\begin{center}
\vskip 0cm
\hspace*{0mm}
\includegraphics[width=65mm,height=65mm]{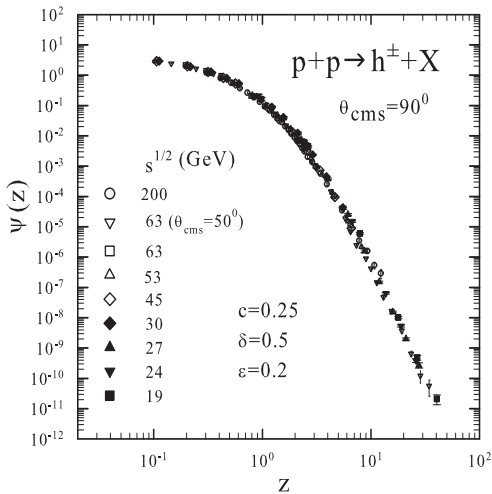}
\vskip -0mm
\caption{
The scaling function $\psi(z)$
for charged hadron production in $pp$ collisions
at the energy $\sqrt s =19-200$~GeV and $\theta_{cms}=90^0$ \protect\cite{PRD75}.
}
\end{center}
\end{figure}
It demonstrates the energy independence of the scaling function $\psi(z)$
of charged hadron production in $pp$ collisions at $\sqrt s =19-200$~GeV,
transverse momenta $p_T=$ 0.2-10~GeV/{\it c}, and $\theta_{cms}=90^0$.
As seen from the figure the function $\psi(z)$ changes
within the range of 11 orders of magnitudes.
The result manifests applicability of the $z$-scaling
over a wide kinematical range.

\section{New LHC data vs. saturation of $\psi(z)$ at low $z$}


The new data on charged hadrons production measured by the CMS Collaboration at LHC confirm
$z$-scaling for $z\le 3$ at midrapidity.

Figure 3(a) shows the charged hadron spectra in $pp$ collisions at
$\sqrt {s_{NN}} = 19-2360$~GeV and  $\theta_{cms}\simeq 90^0$
as a function of transverse momentum $p_T$.
The results are shown for the minimum bias events.
The distributions are measured in the momentum range
 $0.15<p_T<10$~GeV/c.
As seen from Fig. 3(a) the spectra demonstrate strong
dependence on the collision energy.
The ratio of yields at RHIC and ISR increases
with the transverse momentum.
It is of the order of $10^{4}$ at $p_T\simeq 6$~GeV/c.
The shape of the spectra is characterized by
the power behavior at $p_T>1.$~GeV/c.
The CMS data \cite{CMS_pp} reveal similar tendencies as data  at lower energies.
\begin{figure}
\begin{center}
\vskip 0cm
\hspace*{0mm}
\includegraphics[width=65mm,height=65mm]{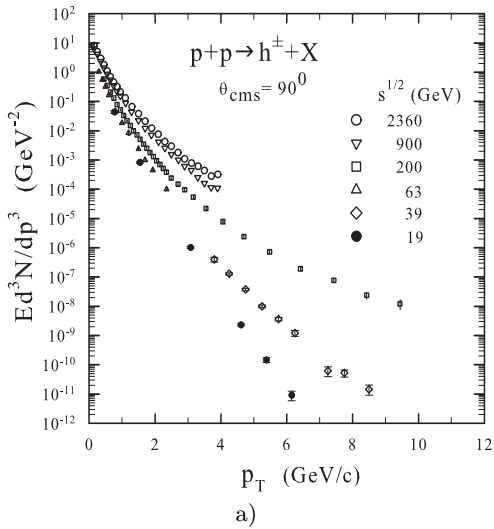}

   a)
\vskip 5mm
\includegraphics[width=67mm,height=67mm]{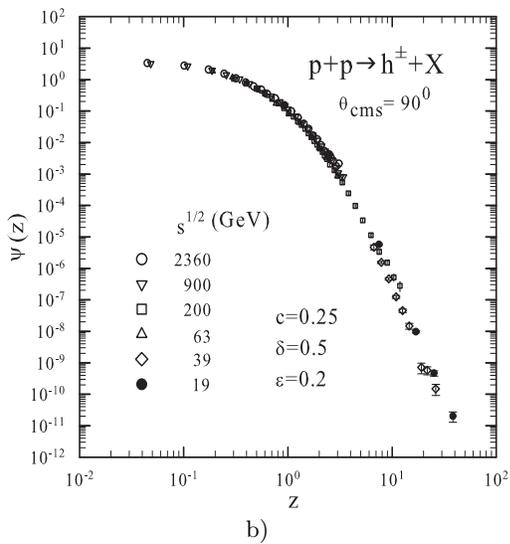}
\vskip -0mm

 b)
\caption{(a) The spectra of charged hadrons produced
 in $pp$ collisions
at the energy $\sqrt s =19-2360$~GeV and $\theta_{cms}=90^0$
in $p_T$-presentation.
(b) The corresponding scaling function $\psi(z)$. The experimental data
 are taken from \protect\cite{ISR_pp1,Cronin_pp,Jaffe_pp,STAR_pp,CMS_pp}.}
 \end{center}
\end{figure}

Figure 3(b) demonstrates $z$-presentation
of the same spectra \cite{ISR_pp1,Cronin_pp,Jaffe_pp,STAR_pp,CMS_pp}.
As seen from the figure the first LHC data confirm the energy independence
of the scaling function with the same values of the
parameters $\delta, \epsilon$ and $c$.
The universal shape of the scaling function is examined in the new energy range.
Evidence of saturation of $\psi(z)$ at low $z<0.2$ is observed.
The minimal values of $z$=0.2 and  $z$=0.045 at $\sqrt s =63$~GeV and 2360~GeV correspond
to $p_T=0.30$~GeV/c and $p_T=0.15$~GeV/c, respectively.
At higher $z>2$ indication on the power behavior
$\psi(z)\sim z^{-\beta}$ is seen.

The behavior of $\psi(z)$ at still lower $z$ can be investigated
by increasing the collision energy $\sqrt s $,
multiplicity density $dN_{ch}/d\eta$ or by decreasing the transverse momentum $p_T$.
For minimum bias events, the multiplicity density at $\sqrt s =7$ and 14~TeV
is estimated to be 5.5 and 6.36, respectively.
Using Eq. (\ref{eq:r9}) for $p_T=0.15$~GeV/c ($\Omega\simeq 1$), it
gives $z$=0.0419 and $z$=0.042 for charged hadrons produced at these energies
at midrapidity. The increase of energy does not change the values of $z$
too much in this region.

Special selection of events with a high multiplicity density
($dN/d\eta\simeq 20$) allows us to reach smaller value of $z\simeq$ 0.03 at $p_T=0.15$~GeV/c.
The value is however still much higher than $z\simeq 10^{-3}$ achieved
at the ISR and RHIC for the identified hadrons \cite{IJMP24}.
Therefore study of production of hadrons with heavy flavors such as $J/\psi$ or $\Upsilon$ is
more preferable for verification of the saturation of $\psi(z)$.
It allows to reach the values $z\simeq 10^{-3}$ at $\sqrt s = 7$~TeV
for $p_T=0.15$~GeV/c.
It was assumed in \cite{IJMP24} that the asymptotic behavior
of $\psi (z)$ at $z\rightarrow 0$ reflects general properties
of the produced system consisting of many constituents.
The universal scaling behavior in this region suggests that
mechanism of particle production at low $p_T$ is governed
by soft self-similar processes which reveal some kind
of a mutual equilibrium leading to the observed saturation.

We expect that verification of the saturation of $\psi(z)$
and/or search for violation of the $z$-scaling at low $z$ could give
new information on the non-perturbative regime and phase transitions
in non-Abelian theories.
It is assumed that discontinuity of the single parameter $c$ describing
the behavior of the scaling function in this region could be a signature
of the Critical Point. Search for the location of CP on the QCD phase diagram
is the main task of the Beam Energy Scan program
at the SPS \cite{BES_SPS} and RHIC \cite{BES_RHIC}.
The scaling assumption makes the CP search relevant for the energies achieved
at the LHC as well.

We would like to emphasize that
"scaling" and "universality" are the concepts
developed to understand critical phenomena.
Scaling means that systems near the critical points
exhibiting self-similar properties are invariant under
transformation of scale. According to universality,
quite different systems behave in a remarkably
similar fashion near the respective critical
points \cite{Stanley}. We see that the $z$-scaling
possesses saturation and universality as remarkable
 properties both in the low $z$ region.

\begin{figure}
\begin{center}
\vskip 0cm
\hspace*{0mm}
\includegraphics[width=65mm,height=65mm]{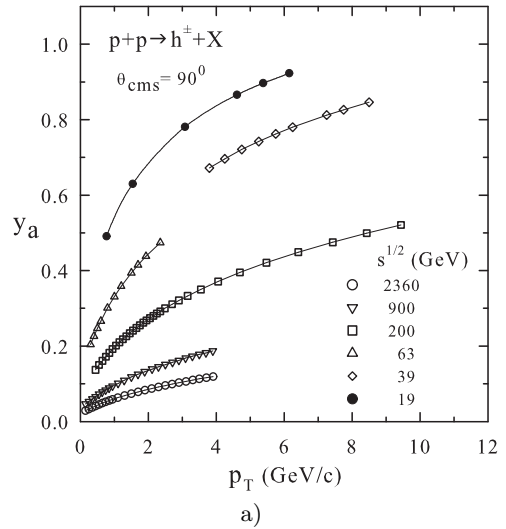}

 a) 
 \vskip 5mm
\includegraphics[width=65mm,height=65mm]{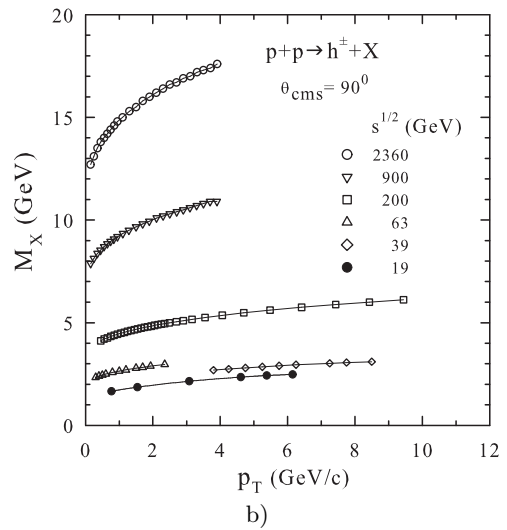}

 b)
\caption{The momentum fraction $y_a$ (a) and recoil mass $M_X$ (b)
for charged hadron production in $pp$ collisions
at the energy $\sqrt s =19-2360$~GeV and
 $\theta_{cms}=90^0$ as~a~function~of~$p_T$.}
\end{center}
\end{figure}

\section{Self-similarity \& Energy loss }

The measured spectrum at the highest collision energy
$\sqrt {s} = 2360$~GeV (Fig.3)
allows us to estimate the constituent energy loss
for charged hadron production in $pp$ collisions
and compare it with the similar one at lower energies
$\sqrt {s} = 19-200$~GeV.
The energy dissipation is proportional to the value of $(1-y_a)$.

Figure 4(a) shows the dependence of the fraction
$y_a$ on the transverse momentum at $\sqrt s = 19-2360$~GeV.
The behavior of $y_a$ demonstrates a monotonic growth with $p_T$.
It means that the energy loss associated with the production of
a high-$p_T$ hadron is smaller than for hadrons
with lower transverse momenta.
The largest energy loss corresponds to the region of low-$p_T$.
The decrease of $y_a$ with the collision energy
represents larger energy losses for higher collision energies.
For $p_T\simeq 4$~GeV/c, the energy loss is estimated to be
about 20\% at $\sqrt s = 19$~GeV and
about 90\%  at $\sqrt s = 2360$~GeV,   respectively.
The study of evolution of the energy loss with the collision
energy has relevance to the evolution of the created nuclear
matter and can be useful for searching for signatures
of the phase transition and the Critical Point in the region of small $p_T$.

The energy dissipation in the final state is connected with the recoil mass $M_X$.
This is another characteristic \cite{PRD75,IJMP24} of the constituent interactions.
It is function of the  momentum fractions
$x_1$ and $x_2$ of the interacting
objects with the masses $M_1$ and $M_2$ and the fraction $y_b$
of the produced recoil object carried by the mass $m_2$.
The recoil mass $M_X$  depends on the constituent interaction
and is connected with the processes of formation of the individual hadrons.

Figure 4(b) demonstrates the dependence of the recoil mass $M_X$ on the transverse momenta
of the charged hadrons  produced in $pp$ collisions
at the energy $\sqrt s = 19-2360$~GeV in the central rapidity region.
The curves  for $\sqrt s = 19, 63$, and 200~GeV demonstrate
slow growth with $p_T$ followed by a successive flattening.
The recoil mass at LHC energies is considerably larger
than at RHIC and SPS energies.
For $p_T\simeq 4$~GeV/c it was found to be about $M_X=18$~GeV at
$\sqrt s =2360$~GeV which is much higher than the corresponding value
$M_X=2$~GeV at $\sqrt s =19$~GeV.
The large recoil mass
means that the momentum of the inclusive particle
is compensated with the momentum of a high multiplicity system
consisting of more particles.

\section{Conclusions}

We have presented results of our analysis
of the first data \cite{CMS_pp} on inclusive spectra of charged hadron
production in $pp$ collisions at the energy $\sqrt s=900$
and 2360~GeV obtained by the CMS Collaboration at the LHC.
The transverse momentum spectra in $p_T$- and $z$-presentation
are compared with data obtained at lower energies
at ISR, S$\bar{\rm p}$pS, SPS, and RHIC.

Based on the results presented here we conclude
that the first LHC data on charged hadron production
in $pp$ collision confirm $z$-scaling.
The energy independence of the scaling function  $\psi(z)$
at the LHC energies in midrapidity range is observed.
Results of the analysis of the CMS data support assumption
on saturation of $\psi(z)$ at low $z$ down to $z\simeq $0.05.
The constituent energy loss and recoil mass $M_X$ at the LHC energies
were estimated as functions of the transverse momentum
in the $z$-scaling approach.
The energy loss increases with  $\sqrt s$ and decreases with increasing $p_T$.
The recoil mass $M_X$ increases with $\sqrt s $ and $p_T$.

We consider that the Beam Energy Scan program in $pp$
collisions at the LHC could be of interest for searching
for scaling violation, phase transition, and location of the Critical Point
in the new energy region.

{
The investigations have been
supported by the IRP AVOZ10480505
and by the special program of the
Ministry of Science and Education of the Russian Federation,
grant RNP.2.1.1.2512.}



\begin{thebibliography}{99}

\bibitem{CMS1}
CMS Collaboration, "The CMS experiment at the CERN LHC"
2008 {\it  JINST} {\bf 3} S08004.
doi:10.1088/1748-0221/3/08/S08004.
 \bibitem{CMS2}
The CMS Collaboration,
"CMS Physics Technical Design Report",
Volume II: Physics Performance
2007 {\it J Phys G: Nucl. Part. Phys.} {\bf 34} 995.
doi:10.1088/0954-3899/34/6/S01.
\bibitem{CMS3}
The CMS Collaboration,
"CMS Physics Technical Design Report: Addendum on
High Density QCD with Heavy Ions",
2007 {\it J. Phys. G: Nucl. Part. Phys.} {\bf 34} 2307.
doi:10.1088/0954-3899/34/11/008.

\bibitem{ATLAS1} ATLAS Collaboration,
The ATLAS Experiment at the CERN Large Hadron Collider. By ATLAS Collaboration
(Bentvelsen S et al.) 2008 {\it JINST} {\bf 3} S08003.
\bibitem{ATLAS2}
The ATLAS Collaboration,
 Expected Performance of the ATLAS Experiment. Detector, Trigger and Physics. CERN-OPEN-2008-020, Geneva, December 2008.

\bibitem{ALICE1} ALICE Collaboration, ALICE: physics performance report, volume I,
2004 {\it J. Phys.} G  {\bf 30} 1517.
\bibitem{ALICE2}
ALICE Collaboration, ALICE: physics performance report, volume II,
2006 {\it J. Phys.} G {\bf 32} 1295.




\bibitem{ISR_pp1} Alper B  et al. (BS Collaboration),
1975 {\it Nucl. Phys.} B {\bf 100} 237.
\bibitem{ISR_pp2}
Drijard  D et al. (CDHW Collaboration),
1982 {\it Nucl. Phys.} B {\bf 208} 1.

\bibitem{Cronin_pp} Antreasyan  D et al. 1979 {\it Phys. Rev.}  D {\bf 19} 764 .

\bibitem{Jaffe_pp}  Jaffe  D E et al. 1989 {\it Phys. Rev.} D {\bf 40}  2777.

\bibitem{STAR_pp} Adams J  et al. (STAR Collaboration), 2003
{\it Phys. Rev. Lett.} {\bf 91}, 172302.


\bibitem{Bjorken}
Bjorken J D  1969 {\it Phys. Rev.} {\bf 179} 1547;
Bjorken J D and Paschos  E A 1969 {\it Phys. Rev.} {\bf 185} 1975.

\bibitem{Feynman}
Feynman R P 1969 {\it  Phys. Rev. Lett.} {\bf 23} 1415 .

\bibitem{Pol}
Polyakov A M 1970  {\it  Zh. Eksp. Theor. Fiz.} {\bf 59} 542;
{\it  Zh. Eksp. Theor. Fiz.} {\bf 60} 1572.

\bibitem{KNO}
Koba Z, Nielsen H B and Olesen P 1972 {\it Nucl. Phys.}  B {\bf 40} 317.


\bibitem{Matveev}
   Matveev V A, Muradyan R M and Tavkhelidze A N
 1971 {\it Part. Nuclei } {\bf 2} 7;
  1972 {\it Lett. Nuovo Cim.} {\bf 5}  907;
1973 {\it Lett. Nuovo Cim. } {\bf 7} 719.

\bibitem{Brodsky}
 Brodsky  S and  Farrar G
 1973 {\it Phys. Rev. Lett.} {\bf 31} 1153;
 1975 {\it Phys. Rev.}  D {\bf 11} 1309.

\bibitem{Bialas} Bialas A and  Peschanski R
1986  {\it Nucl. Phys.} B {\bf 273} 703;
1988  {\it Nucl. Phys.} B {\bf 308} 857.

\bibitem{Dremin}  Dremin I M  1987 {\it JETP Lett.} {\bf 45} 643.

\bibitem{DeWolf} DeWolf E A, Dremin I M  and Kittel W
1996 {\it Phys. Rep.} {\bf 270} 1.

\bibitem{Z1}
Zborovsk\'{y} I, Panebratsev Yu, Tokarev M and  \v{S}koro G
1996 {\it Phys. Rev.} D {\bf 54} 5548.
\bibitem{Z2}
 Tokarev M, Zborovsk\'{y} I, Panebratsev Yu and  \v{S}koro G
2001 {\it Int. J. Mod. Phys.} A {\bf 16}  1281.

\bibitem{PRD75}
Zborovsk\'{y} I and Tokarev M V  2007 {\it Phys. Rev.} D {\bf 75} 094008.

\bibitem{IJMP24}
Zborovsk\'{y} I and Tokarev M V  2009 {\it Int. J. Mod. Phys.} A {\bf 24} 1417.


\bibitem{CMS_pp}
CMS Collaboration, "Transverse-momentum and pseudorapidity distributions of
charged hadrons in $pp$ collisions at $\sqrt s = 0.9$ and 2.36~TeV",
http://www.arxiv.org/abs/1002.0621v2
2010 {\it JHEP} {\bf 02} 041.

\bibitem{BES_SPS}
NA61 Collaboration,
Study of Hadron Production in Hadron-Nucleus and Nucleus-Nucleus
Collisions at the CERN SPS.
CERN SPSC-2006-034, SPSC-P-330, November 3, 2006.



\bibitem{BES_RHIC}
Caines H (for the STAR Collaboration), Proceedings for the Rencontres de Moriond 2009 QCD session, [arXiv:0906.0305v1].\\
Abelev B I  et al., [STAR Collaboration] SN0493: Experimental Study of the QCD Phase Diagram and Search for the Critical Point: Selected Arguments for the Run-10 Beam Energy Scan, http://drupal.star.bnl.gov/STAR/starnotes/public/sn0493.

\bibitem{Stanley}
 Stanley H E
 1999 {\it  Rev. Mod. Phys.} { \bf 71}, S358.

\end{thebibliography}
\end{document}